\documentclass{IEEEcsmag}
\usepackage[hyphens]{url}
\usepackage[utf8]{inputenc}
\usepackage[colorlinks,urlcolor=blue,linkcolor=blue,citecolor=blue]{hyperref}
\usepackage{lettrine}
\usepackage{comment}
\usepackage[table,xcdraw]{xcolor}

\newcommand{\jnote}[1]{\textcolor{blue}{JEFF: #1}}

\newcommand\blfootnote[1]{%
  \begingroup
  \renewcommand\thefootnote{}\footnote{#1}%
  \addtocounter{footnote}{-1}%
  \endgroup
}

\jvol{XX}
\jnum{XX}
\paper{8}
\jmonth{Mon}
\jname{Computing in Science \& Engineering}
\pubyear{2022}

\title{A Secure Future for Open-Source Computational Science and Engineering}
\author{Reed Milewicz}
\author{Jeffrey Carver}
\author{Samuel Grayson}
\author{Travis Atkison}
\date{July 2022}

\begin{document}

\maketitle

%%%%%%%%%
% Potential Outline For Paper (adapted from the current text)
%
% I. What is the problem?
% II. Overview of the attack vector presented by research software
% III. Specific aspects of the research software environment that contribute to the risk:
%   - Software
%   - Data
%   - Compute environments
% IV. Examples of previous attacks
%   - Attacks on data
%   - Attacks on compute
% V. Potential for future attacks
%   - Attacks on software
%   - Attacks on data
%   - Attacks on compute
% VI. Proposed Solution
%   - Existing tools
%   - New research needed
%
% Need to see if we can run the theme of software, data, compute throughout the second half of the article as a theme.

\lettrine[lines=3]{J}{ournalists}, public policy analysts, and economists have called attention to the growing importance that high-performance and scientific computing have to national security and industrial leadership. 
As computing continues to power scientific advances in virtually every discipline, so too does it improve our economic productivity and quality of life. 
In 2020, a US government-funded study by Hyperion Research estimated that every dollar invested into HPC resulted in an average of \$507 in new revenue and \$47 in profit or cost-savings~\cite{hyperion2020update}. 

The increasing social, political, and economic importance of research software has also brought the question of software security to the fore. 
Just as unintentional software errors can threaten the integrity of scientific studies, malicious actors could leverage vulnerabilities to alter results, exfiltrate data, and sabotage computing resources. 
A 2015 US Department of Energy workshop report on cybersecurity in scientific computing vividly captured these concerns, arguing that attacks on scientific computing could inject flaws into critical infrastructure, inhibit public policy decision-making, and weaken US economic competitiveness~\cite{peisert2015ascr}.
With concerns that the world is returning to an era of great power competition, these threats could ultimately harm the ability of researchers to collaborate openly, curtailing the free flow of ideas that is key to innovation. 
Likewise, a climate of uncertainty around scientific computing could amplify public distrust in scientific institutions.

For these reasons, research has been proceeding to develop solutions to detect, mitigate, and prevent security threats at the point of deployment on HPC systems, such as through hardware-based trusted execution environments, access controls, event logging, and secure data provenance~\cite{li2022cybersecurity}. 
While these are encouraging advances, we must also look to addressing security earlier in the research software development lifecycle. 
As software security expert Rod Cope has noted, ``While there is---rightly---a big focus on securing software that is already deployed, the reality is that many future vulnerabilities stem from the creation of that software\ldots\ Not securing software right from the development stage is like putting a deadbolt on a cardboard door''~\cite{cope2020strong}. 

This emphasis on integrating security as early as possible in the development process is representative of the ``shift left'' philosophy of the DevSecOps movement which has recently gained traction in industry\footnote{\url{https://www.devsecops.org}}. In this new line of thinking, security advocates argue that it is critical to understand where and how software is created so security can be moved closer to that development --- that is, earlier in the software lifecycle and with developers engaging with security in their work.

Increasingly, the development of research software occurs on the open web. 
In the past decade, research software development has transitioned from independently developed research scripts to community-driven, open-source software that leverages modern software development technologies. Today, even a ``bare-bones'' modeling and simulation code may rely on different open-source packages for I/O, parallel communication, meshing, discretization, solvers, and visualization. The code teams responsible for these packages belong to many different institutions and can vary widely in terms of their size and maturity. In short, the leading edge of research software development is increasingly large-scale, open-source, and distributed across teams with different values and perspectives on software development practice.

While these changes have accelerated the pace of progress, the shift towards an interconnected research software ecosystem has also complicated security considerations with a proliferation of environments, actors, and attack surfaces.
In terms of people, the production of research software can involve many teams of researchers and research software engineers, rotating casts of student interns and postdocs, and academic and industry partners. 
In terms of infrastructure, the software makes its way through numerous environments, from open-source hosting and continuous integration platforms to national labs and university servers and finally to clusters and HPC facilities for code execution.
In terms of data, the large amounts of data produced by the software must move to various types of storage for analysis and reuse. 
A compromise of the actors, infrastructure, or data could have disastrous consequences.

\vspace{8pt}

\lettrine[lines=3]{M}{ost} reported attacks on research computing institutions so far appear to be financially motivated, where the attackers tried to extort ransoms, hijack compute resources, or exfiltrate data.
%Ransomware attacks, where hackers seize control of valuable digital assets and threaten to release or destroy them unless their demands are met, have increased dramatically in recent years. These attacks are spilling over into academia. 
For example, in May 2020, hackers leveraged a vulnerability in the VPN at Michigan State University to carry out a ransomware attack against the Physics and Astronomy Department. 
Between 50 and 70\% of research activities were temporarily halted due to the attack, and some research groups lost years' worth of data~\cite{adams2021research}. 
There have been other similar ransomware-style attacks recently reported in the media that targeted the University of Amsterdam, Amsterdam University of Applied Sciences, the UK Research and Innovation Agency, Maastricht University, and the Netherlands Organization for Scientific Research~\cite{enserink2021ransomware}. 
In another type of attack, supercomputing centers in the UK, Germany, Switzerland, and Spain fell victim to cryptocurrency-mining hijackers~\cite{cimpanu2020supercomputers}. 
In the third type of attack, security analysts at ESET identified Kobalos, a previously unknown malware that targets HPC clusters and university servers by creating backdoors for hackers to steal passwords~\cite{leveille2021kobalos}. 
The sophistication and targeted nature of these types of malware suggests that malicious actors may be plotting focused attacks on research computing infrastructure.

At the time of writing, there have been no reported attacks using research software, systems, or data as the principal attack vector. 
However, as researchers increasingly develop and use open-source software there is likely to be a consequent increase in attacks on research software.
A 2021 report by analysts at Sonatype showed a 650\% year-over-year increase in attacks on open-source software supply chains~\cite{helpnet2021opensource}. 
The ubiquity of open-source software packages means that these attacks can be far-reaching.
For example, the US government estimated a vulnerability disclosed late last year in Log4J, a popular open-source infrastructure package, potentially affects hundreds of millions of devices~\cite{barr2022log4j}. 
Security experts anticipate open-source software will be the next frontier for cyberattacks~\cite{carder2022ossattacks}. 

Within the research software domain, there are multiple approaches attackers could use to introduce malicious code or novel vulnerabilities. 
First, attackers could hijack a trusted contributor's account (e.g. on GitHub) and modify source code hosted in an online repository. Second, attackers could create a seemingly useful package with hidden malicious code, perhaps typo-squatting on a well-known package.
%Second, attackers could create a seemingly useful package with hidden malicious code, perhaps typo-squatting on a well-known package (e.g. `urlib3' was created to spoof `urllib3' \cite{vu2020typosquatting}).
Third, an attacker could tamper with continuous integration or deployment services (e.g., TravisCI or Jenkins) and quietly alter software prior to deployment. The recent spate of cryptomining and ransomware attacks shows these types of attacks are already possible.
%Finally, if an attacker manages to gain privileged access to production environments, they could alter source code and binaries even more surreptitiously. \sam{\textit{Is this paragraph about how attackers get in, or what they do once they are in? If the former, strike this point.}}

Meanwhile, if an attacker can infiltrate a critical piece of a research software stack, they have many options available to create mischief. 
While a financially motivated attacker might commit the attacks seen above, an attacker motivated by attacking research sensitive to national security could be even more dangerous:
\begin{itemize}
\item They could introduce code to intentionally produce inaccurate results that could lead to sub-optimal decisions or faulty science. 
\item They could exfiltrate confidential research data. 
\item They could sabotage the compute infrastructure by introducing code to uselessly consume compute cycles or clog disk drives with garbage.
\end{itemize}

It is even more concerning considering that all of these threats can be multiplied across the open-source software ecosystem. Each third-party dependency a scientific codebase has could have vastly different coding practices and attack surfaces.
If \textit{any one} of those dependencies is vulnerable, the whole system that uses it is vulnerable.

\vspace{8pt}

\lettrine[lines=3]{T}{he} spectre of cyberattacks on open-source research software is not just a material threat but also an epistemic one: it endangers our ability to trust our codes and the resulting science. 
As such, we need to build security into scientific computing from the ground up, starting with research software development. 
Shifting security left, however, means confronting the fact that research software developers are not cybersecurity experts or vice versa. 
Not only do most researchers lack knowledge of security, but the typical cybersecurity analyst is also wholly unfamiliar with the languages, libraries, and programming models of scientific computing.
This disconnect leaves open questions about how to move forward with incorporating cybersecurity principles into the research software development process.

In recent years the rifts between scientific and mainstream software engineering have begun to close with the rise of the research software engineer (RSE) professional identity and the emergence of community organizations advocating for SE best practices (e.g. Better Scientific Software and the US-RSE Association).
As a result, a cross-disciplinary movement has emerged with the goal of translating insights from mainstream software engineering and to pioneer new tools and practices well-suited to the scientific domain. 

We see a similar opportunity for addressing the disconnect between cybersecurity requirements and research software development needs.
By incorporating cybersecurity principles throughout the research software lifecycle, especially early in the process, teams can begin to address the concerns raised above.
The Confidentiality Integrity and Availability (CIA) triad is a standard model for designing secure information systems. 
The CIA triad defines the necessary characteristics for a system to be secure. 
\textit{Confidentiality} is the securing of private information from unauthorized entities. 
\textit{Integrity} is the preservation of the original message without any alterations. 
\textit{Availability} is the accessibility to an entity. 
Table~\ref{table:cia} lists examples of the types of attacks that fall into each of these categories.
If the information system can meet appropriate levels of these requirements, it meets an acceptable threshold of security.

\begin{table}[!htb]
\caption{Example attacks for each CIA category}
\label{table:cia}
\resizebox{\columnwidth}{!}{%
\begin{tabular}{ccc}
\rowcolor[HTML]{3166FF} 
{\color[HTML]{FFFFFF} \textbf{Confidentiality}} & {\color[HTML]{FFFFFF} \textbf{Integrity}} & {\color[HTML]{FFFFFF} \textbf{Availability}} \\
Man-in-the-Middle & Masquerade & Jamming \\
Eavesdropping & Code Injection & DOS \\
Side-Channel & Code Modification & Delay of Information
\end{tabular}
}
\end{table}

We identify two types of steps to take moving forward: (1) porting and adopting existing cybersecurity practices into the research software domain and (2) conducting research on how those cybersecurity practices and others work in the research software domain.

First, well-known, proven cybersecurity practices can be integrated into the research software development process. In general, these practices fall under the concept of ``building security in'' rather than trying to bolt it on at the end. These practices include:
\begin{itemize}
    \item Considering cybersecurity in the requirements specification step
    \item Architecting and designing software with cybersecurity considerations in mind
    \item Using ``safe'' coding practices to reduce the potential of introducing exploitable vulnerabilities
    \item Incorporating security checks into build and test automation to identify problems prior to release    
    \item Using vulnerability scanners to identify problematic code
\end{itemize}

Second, additional research into the adoption of cybersecurity practices in research software is needed.
The research agenda includes answering questions like:
    \begin{itemize}
        \item How do existing cybersecurity tools and practices need to be modified to work in a research software ecosystem that is dominated by research software developers and RSEs?
        \item How do we encourage the adoption of cybersecurity tools and practices among scientific computing professionals? What can we do to lower the barriers to entry and increase security literacy?
        \item What implications does the use of cybersecurity tools and practices have on the performance, reproducibility, and openness of research software?
        \item What are the right criteria by which we can judge the overall success of security tool and practice use? That is, when will we know that we have succeeded?
    \end{itemize}

In the long run, we believe cybersecurity could eventually become ``nativized'' into research software development praxis in the same way that software engineering concepts and tools have been adopted. 
Many security principles can, with some reframing, fit into what researchers already value, such as the FAIR principles for research software~\cite{chue_hong_neil_p_2022_6623556}.
\begin{itemize}
    \item \textbf{Findable} software means the software has a globally unique identifier (such as a DOI or Software Heritage ID). This facilitates the creation of ``manifest'' files that precisely specify dependencies. Manifest files improve security because they can be automatically audited for recency and known vulnerabilities.
    \item \textbf{Available} software means it is retrievable by that identifier from known repositories. This practice improves the security of the software supply chain, since the dependencies come from reliable, trusted sources over an encrypted channel.
    \item \textbf{Interoperable} software means that software operates behind explicit interfaces that can be incorporated in a wide range of applications, rather than each author writing a library for their own use-case. This improves security by reducing the attack surface and concentrating efforts on improving a small set of core libraries.
    \item \textbf{Reusable} software means the software has the metadata necessary to reuse software results (provenance) and the ability to get the same results in a reuse (reproducibility). Reproducibility helps researchers detect tampering, since they can easily compare a hash of the results, while provenance can help researchers debug security incidents when they occur.
\end{itemize}

Our hope is that researchers may become conversant in security and able to work with professionals in cybersecurity and software engineering to achieve it in practice.

\blfootnote{Sandia National Laboratories is a multimission laboratory managed and operated by National Technology \& Engineering Solutions of Sandia, LLC, a wholly owned subsidiary of Honeywell International Inc., for the U.S. Department of Energy’s National Nuclear Security Administration under contract DE-NA0003525. Any subjective views or opinions that might be expressed in the paper do not necessarily represent the views of the U.S. Department of Energy or the United States Government. SAND2022-10431-O.}

\bibliographystyle{IEEEtran}
\bibliography{bibliography}

\begin{IEEEbiography}{Jeffrey Carver}{\,} Ph.D. is a full professor in the Department of Computer Science at The University of Alabama. His general research interest lie in the area of empirical software engineering and human factors in software engineering. His research focuses on topics including: software engineering for research software, software quality assurance, peer code review, software security, human errors, and software engineering education.
\end{IEEEbiography}

\begin{IEEEbiography}{Reed Milewicz}{\,} Ph.D. is a computer scientist and software engineering researcher in the Department of Software Engineering and Research at Sandia National Laboratories. His research focuses on developing better practices, processes, and tools to improve software development in the scientific domain. He leads software science research efforts within his department and is a member of the Interoperable Design of Extreme-Scale Application Software (IDEAS-ECP) project, an arm of the Exascale Computing Project, where he is part of the Productivity and Sustainability Improvement Planning (PSIP) team.
\end{IEEEbiography}

\begin{IEEEbiography}{Samuel Grayson}{\,} is a Ph.D. candidate at the University of Illinois at Urbana-Champaign advised by Darko Marinov and Daniel S. Katz. Sam's research focus is on improving the process of developing software for computational sciences using software tools and process improvements. He is also a summer intern at Software Engineering and Research at Sandia National Laboratories.
\end{IEEEbiography}

\begin{IEEEbiography}{Travis Atkison}{\,} Ph.D. is an associate professor in the Department of Computer Science at The University of Alabama.  His current research efforts focus on the topics of cyber security, transportation infrastructure, and control systems security. These efforts include malicious software detection, threat avoidance, digital forensics, and security in control system environments (in both power systems and transportation). 
\end{IEEEbiography}

\end{document}